\documentclass[aps,prl,superscriptaddress,twocolumn,amsmath,amssymb]{revtex4}
\usepackage{graphicx}
\usepackage{epstopdf}
\usepackage{bm}
\usepackage{wasysym}
\begin{document}

\title{Raman Scattering Signatures of Kitaev Spin Liquids in A$_2$IrO$_3$ Iridates}

\author{J.~Knolle}
\affiliation{Max Planck Institute for the Physics of Complex Systems, D-01187 Dresden, Germany}

\author{Gia-Wei Chern}
\affiliation{Center for Nonlinear Studies and Theoretical Division, Los Alamos National Laboratory, Los Alamos, NM 87545, USA}

\author{D.~L.~Kovrizhin}
\affiliation{T.C.M.~Group, Cavendish Laboratory, J.~J.~Thomson Avenue, Cambridge CB3 0HE, United Kingdom,\\
and RRC Kurchatov Institute, 1 Kurchatov Square, Moscow 123182, Russia }

\author{R.~Moessner}
\affiliation{Max Planck Institute for the Physics of Complex Systems, D-01187 Dresden, Germany}

\author{N.~B.~Perkins}
\affiliation{Department of Physics, University of Wisconsin, Madison, Wisconsin 53706, USA}

\begin{abstract}
We study theoretically the Raman scattering response $I(\omega)$ in the gapless quantum spin liquid phase of the Kitaev-Heisenberg model.
The dominant polarization-independent contribution $I_K(\omega)$ reflects the density of states of the emergent Majorana fermions in the ground-state flux-sector. The integrability-breaking Heisenberg exchange generates a second contribution, whose dominant part $I_H(\omega)$ has the form of a quantum quench corresponding to an abrupt insertion of four $Z_2$ gauge fluxes. 
This results in a weakly polarization dependent response with a sharp peak at the energy of the flux excitation accompanied by broad features, which can be related to Majorana fermions in the presence of the perturbed gauge field. We discuss the experimental situation and explore more generally the influence of integrability breaking for Kitaev spin liquid response functions.
\end{abstract}

\maketitle

{\it Introduction.}
Frustrated magnetic materials hold the promise of showing a wide variety of novel cooperative quantum phenomena. Frustration can arise when interactions are incompatible with the geometry of the underlying lattice, or as a result of competing interactions. An example of the latter is given by the celebrated Kitaev model \cite{kitaev06}, where half-integer spins, arranged on a honeycomb lattice, interact via anisotropic Ising exchange. The model harbours distinct topologically ordered states, including gapless and gapped quantum spin liquids (QSL). Only a few models up to date are known to exhibit quantum spin-liquid states, and among those the Kitaev model stands out as it offers an exact solution in two dimensions~\cite{kitaev06}. The model serves as a representative of a large class of spin-liquids with Majorana fermions coupled to a $Z_2$ gauge-field. A number of integrable generalizations of the model exist with spin-disordered ground states~\cite{Yao2009,Qi2014,Vaezi2014}, as well as three-dimensional analogies~\cite{Mandal2009,Subhro2014,Kimchi2014,Hermanns2014,Takagi2014}.

Because of the simple form of the Kitaev Hamiltonian there is hope that its physics can be realized in nature. A number of proposals suggest that materials with strong spin-orbit coupling, such as $A_2$IrO$_3$ where $A = $ Na, Li, are possible candidates~\cite{jackeli09,cha10,Singh2012}. In these compounds, Ir$^{4+}$ ions form  weakly coupled hexagonal layers. Due to spin-orbit coupling, the atomic ground state of Ir$^{4+}$ ion is a Kramers doublet in which  spin and orbital angular momentum are entangled. The interactions between these magnetic moments are modeled by the Kitaev-Heisenberg (KH) model, which contains both the anisotropic ferromagnetic Kitaev interaction $J_K$, as well as the isotropic antiferromagnetic Heisenberg exchange $J_H$. Recent theoretical studies indicate that the QSL phase of the Kitaev model is stable with respect to small Heisenberg perturbations, \cite{cha10,Reuther2011,Schaffer12} and thus might be realized in $A_2$IrO$_3$ systems~\cite{Gedik2014}.

Raman scattering is a valuable tool for understanding  antiferromagnetically ordered transition metal oxides because its polarization dependence allows to probe different regions of the Brillouin zone~\cite{devereaux07}, e.g.~in frustrated triangular antiferromagnets~\cite{perkins08,perkins13}, and in high-Tc superconductor parent compounds~\cite{chubukov95,blumberg9697,Devereaux2011,Loidl2013}.
In Mott insulators, the Raman process couples a dynamically induced electron-hole pair with ``two-magnon states'' which reflect the underlying magnetic phase even if a simple spin wave picture of the low energy excitations is not applicable, e.g. in QSLs~\cite{cepas08,ko10}.
Due to the lack of local order a very weak polarization dependence is conjectured to be one of the key signatures of QSLs~\cite{cepas08}. Indeed, recent Raman scattering experiments revealed spin-liquid like features in the Heisenberg spin one-half Kagome-lattice antiferromagnet, herbertsmithite ZnCu$_3$(OH)$_6$Cl$_2$~\cite{wulferding10}.
In addition, given the difficulty of using neutrons to study compounds hosting iridium ions~\cite{Coldea2012},
Raman scattering may be of particular interest for the $A_2$IrO$_3$ series.

Clearly, a detailed quantitative analysis of Raman scattering in a spin liquid is called for, not least because  the experimental task of diagnosing QSLs remains such a challenge. The main result of this work is the identification, in the dynamical Raman scattering response, of signatures of quantum number fractionalization, a hallmark of a topologically ordered phase. We report a theoretical analysis of the inelastic Raman scattering obtained within
the KH model in the limit of small Heisenberg exchange, where the parent ground state is the QSL of the corresponding exactly solvable Kitaev model~\cite{cha10}. 
To leading order in Heisenberg exchange, and neglecting higher order combinations of Majorana fermion density of states, we obtain two dominant contributions to the response $I(\omega )=I_{K}(\omega )+I_{H}(\omega )$,  $I_{K}(\omega )$ originating from the Kitaev term, and $I_{H}(\omega )$ from the Heisenberg perturbation. Our perturbation theory amounts to approximating the ground state of the KH model with the integrable one. However, the calculation of the response goes one step beyond integrability by including contributions to the Raman vertex arising from the integrability-breaking Heisenberg term.

In the remainder of the paper, after introducing the KH model, we derive the expression for the Raman vertex. We then outline the evaluation of $I_K$ and $I_H$, and discuss their salient characteristic features. We close with remarks on the relevance of our results to a broader class of Hamiltonians and observables.

\begin{figure}
\includegraphics[width=0.75\columnwidth]{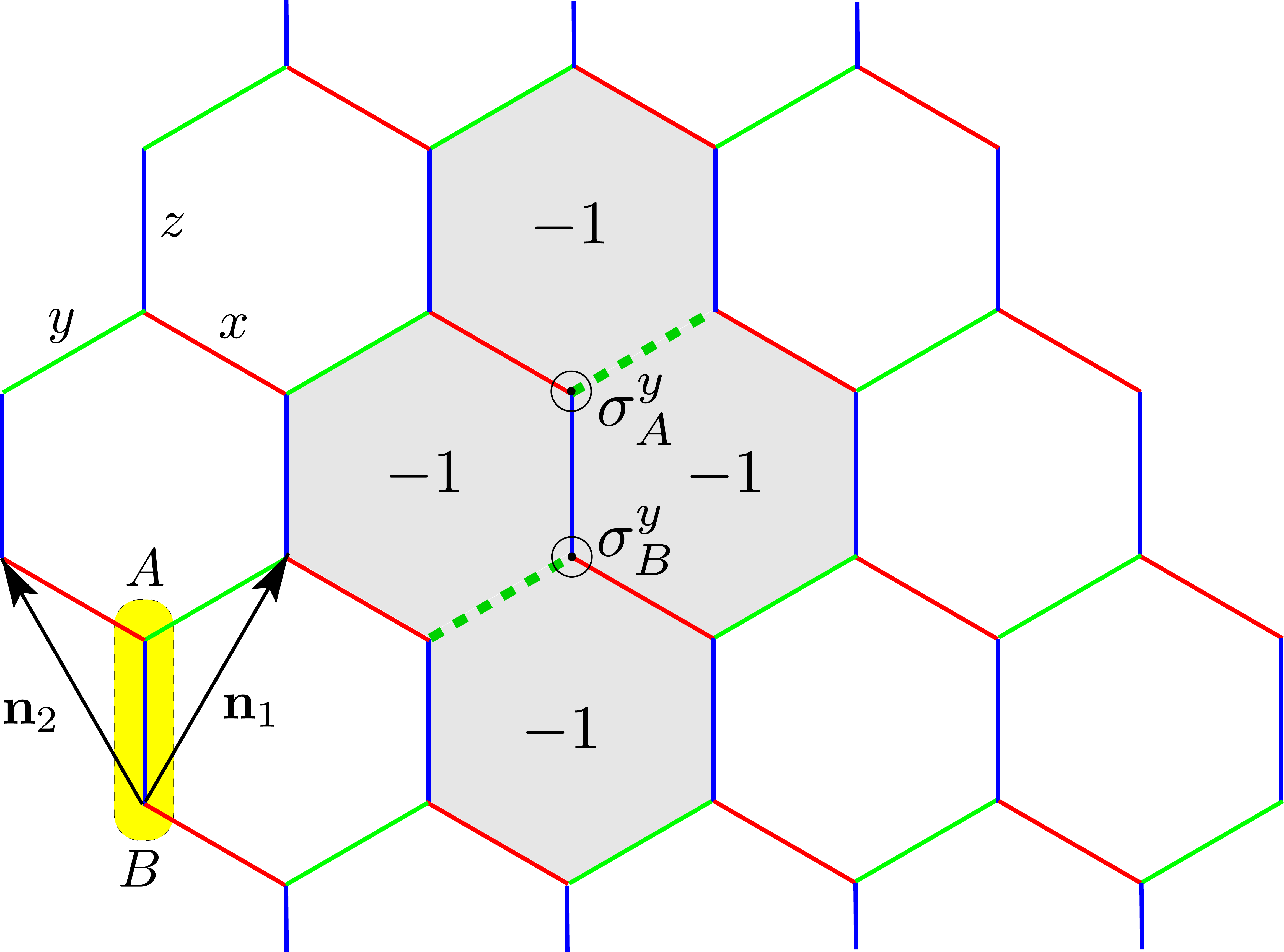}
\caption{\label{fig:honeycomb} Kitaev honeycomb model. The shaded yellow region indicates the unit cell with two sites $A$ and $B$. Three inequivalent nearest-neighbour bonds, denoted as $x,y,z$, are indicated as red, green, and blue correspondingly. The calculation of the Raman response due to the Heisenberg part can be mapped to a local quantum quench, in which four adjoining $Z_2$ fluxes, shown as gray hexagons, are inserted, e.g. the contribution from nearest-neighbor $\sigma_{A\mathbf{r}}^y \sigma_{B\mathbf{r}}^y$ interactions along a $z$-bond which flips the sign of the link variables is shown by green dashed bonds.}
\end{figure}
{\it Model.}
The Hamiltonian of the KH model reads
\begin{eqnarray}
	\label{eq:H1}
	 \mathcal{\hat H} = - J_K \sum_{\langle ij \rangle_a} \hat\sigma_i^{a} \hat\sigma_j^{a}
	+ J_H \sum_{\langle ij \rangle} \hat{\bm \sigma}_i \cdot\hat{\bm \sigma}_j,
\end{eqnarray}
which reduces to the original Kitaev model for $J_H = 0$. As shown in Kitaev's seminal work, the model can be exactly solved in this limit by representing the spin-1/2 operators $\hat\sigma^a_i$  in terms of four Majorana fermion operators $\hat b^x_i, \hat b^y_i,\hat b^z_i$, and $\hat c_i$ such that $\hat\sigma^a_i = i \hat c_i \hat b^a_i$, which satisfy the anti-commutation relations, $\{\hat b^a_i, \hat b^{a'}_j\} = 2 \delta_{ij} \delta_{a, a'}$, $\{\hat c_i, \hat c_j\} = 2\delta_{ij}$, and $\{\hat c_i, \hat b^a_j\} = 0$.
For our purposes it is convenient to introduce complex bond-fermions $\hat \chi^\dagger_{\langle ij \rangle_a} = (\hat b^a_i - i \hat b^a_j)/2$ by combining two ${\hat b}$ Majorana operators on adjacent sites. The Kitaev contribution in Eq.~(\ref{eq:H1}) then takes the form
\begin{eqnarray}
	\label{eq:HK}
	\mathcal{\hat H}_K = i J_K \sum_{\langle ij \rangle_a} \hat u_{\langle ij \rangle_a} \hat c_i\,\hat c_j,
\end{eqnarray}
where bond operators $\hat u_{\langle ij \rangle_a} = i \hat b^a_i \hat b^a_j = 2 \hat \chi^\dagger_{\langle ij \rangle_a} \hat \chi^{\;}_{\langle ij \rangle_a} - 1$ are constants of motion for $\mathcal{\hat{H}}_K$, i.e.~$[\mathcal{\hat{H}}_K, \hat u_{\langle ij \rangle_a}] = 0$.  The Hilbert space in which $\mathcal{\hat H}_K$ acts can now be decomposed into gauge $|F\rangle$ and matter $|M\rangle$ sectors. We denote the ground state of $\mathcal{\hat{H}}_K$ by $|0\rangle = |F_0 \rangle \otimes |M_0 \rangle$, in which $\hat u_{\langle ij \rangle_a} |F_0\rangle = +1 |F_0\rangle$, i.e. we replace the bond operators by their ground-state eigenvalue $+1$. The Kitaev part of the Hamiltonian then reduces to a quadratic form in Majorana fermions ${\hat c}_i$, and thus can be diagonalised. To this end, we first combine two Majorana ${\hat c}$ fermions belonging to two sub-lattices in the unit cell to form a complex fermion (a matter fermion), $\hat f_{\mathbf r} = (\hat c_{A,\mathbf r} + i \hat c_{B, \mathbf r})/2$. After a Fourier transform, followed by a Bogoliubov transformation $\hat f_{\mathbf q} = \cos\theta_{\mathbf q} \hat a_{\mathbf q} + i \sin\theta_{\mathbf q} \hat a^{\dagger}_{-\mathbf q}$, the Hamiltonian of Eq.~(\ref{eq:HK}) in the ground-state flux-sector is diagonalized
$\mathcal{\hat H}_0 = \mathcal{\hat H}_{K, F_0} = \sum_{\mathbf q} | s_{\mathbf q} | (2 \hat a^\dagger_{\mathbf q} \hat a^{\;}_{\mathbf q} - 1 ),$
where $s_{\mathbf q} = J_K (1 + e^{i\mathbf q\cdot\mathbf n_1} + e^{i\mathbf q\cdot\mathbf n_2})$, and $\tan2\theta_{\mathbf q} = -{\rm Im}[s_{\mathbf q}]/{\rm Re}[s_{\mathbf q}]$. The primitive lattice vectors $\mathbf n_1$, $\mathbf n_2$ are defined in Fig.~1. The ground state of the matter sector $|M_0 \rangle$ is defined by the condition that $\hat a_{\mathbf q} | M_0 \rangle  = 0$ for all $\mathbf q$, and the ground-state energy is $E_0 = -\sum_{\mathbf q} | s_{\mathbf q} |$.

We note that the Hamiltonian of Eq.~(2) is defined in the enlarged Hilbert space, and $\mathcal{\hat H}_K$ has a local $Z_2$ gauge invariance. The fermionic spectrum is thus defined by the configurations $\{\phi_{\hexagon}\}$ of $Z_2$ fluxes on hexagons; here $\phi_{\hexagon} = \prod_{\langle ij \rangle \in \hexagon} u_{ij}$ is a product of bond variables. The fermionic ground state lives in the flux-free sector, i.e.~when $\phi_{\hexagon} = +1$ on all hexagons. The physical states $|\Psi_{\rm phys}\rangle=\hat P |\Psi\rangle$ are defined using projector $ \hat P=\frac{1}{2}\hat P'\left[ 1+(-1)^{N_{\chi}}(-1)^{N_{f}}\right]$,
where $\hat P'$ is the sum of all operators that change bond fermion numbers in an inequivalent way~\cite{Yao2009}, and $N_{\chi/f}$ are the bond/matter fermion number operators. Note that for a given state the total parity of $N_{\chi}+N_{f}$ is a conserved quantity being always even, whereas the parity of the corresponding bond/matter sectors can be changed by a gauge transformation. In the remainder we use the property that for a large class of operators, e.g. those that do not change the bond fermion number, matrix elements are the same when calculated in projected or unprojected states.~\cite{Baskaran2007}

Upon addition of small non-zero Heisenberg exchanges to the Hamiltonian $(J_H\neq 0)$, the Kitaev QSL states remain stable \cite{cha10}, with ultra-short-ranged (nearest neighbor only) spin-correlations replaced by exponentially decaying ones~\cite{Schaffer12}. We assume in the following, that $\lambda=J_H/J_K\ll 1$ and take into account the Heisenberg terms perturbatively.

{\it Raman operator.}
We derive the Raman vertex operator along the lines of the Loudon-Fleury approach~\cite{fleuri68,shastry1990}. The former is given by the photon-induced super-exchange, which for the KH model contains two contributions $\mathcal{\hat R} = \mathcal{\hat R}_K + \mathcal{\hat R}_H$ (the counterpart of the Loudon-Fleury vertex for the Heisenberg model)
\begin{eqnarray}
	\mathcal{\hat{R}} = \sum_{\langle ij \rangle_a} (\hat{\bm\epsilon}_{\rm in}\cdot \mathbf d_a)(\hat{\bm\epsilon}_{\rm out}\cdot\mathbf d_a)
	\left(K_K \hat{\sigma}^a_i \hat{\sigma}^a_j + K_H \hat{\bm \sigma}_i \cdot \hat{\bm\sigma}_j\right),
\end{eqnarray}
where $\mathbf d_a$ denote lattice vectors, and $\hat{\bm\epsilon}_{\rm in/out}$ are polarization vectors of the incident/outgoing photons. The constants $K_K \propto J_K$ and $K_H \propto J_H$, hence $\lambda=K_H/K_K\ll 1$.

The Raman response of the KH model~(\ref{eq:H1}) is related to the Fourier transform $I(\omega)=\int_{-\infty}^{\infty} dt e^{i\omega t} i F(t)$ of the correlation function $i F(t) = \langle \mathcal{\hat{R}}(t) \mathcal{\hat{R}}(0) \rangle$, where the average is taken with respect to the ground state $|\Psi_0\rangle$ of the KH Hamiltonian, and the operators $\mathcal{\hat{R}}(t)$ are in their Heisenberg representation. After switching to the interaction representation treating $\mathcal{\hat H}_H$  as the interaction, the correlation function takes the form
\begin{eqnarray}\label{RamanInteractionRep}
	F(t) = -i \langle 0 | \mathrm{T}_K[ \mathcal{\hat{R}}(t) \mathcal{\hat{R}}(0) \,e^{-i \int_{C_K} \mathcal{\hat H}_H(t') dt' } ]| 0 \rangle,
\end{eqnarray}
where time-ordering $\mathrm{T}_K$ and the integral are assumed to be along the Keldysh contour (with the Heisenberg term adiabatically switched on and off at $t\to -\infty $). Note that the average is taken with respect to the ground state $|0\rangle$ of the Kitaev Hamiltonian Eq.~(2). 
Starting from Eq.~(\ref{RamanInteractionRep}) we perturbatively compute the response by expanding
the exponent in powers of $\lambda =J_{H}/J_{K}$, see supplementary
material \cite{suppl}. To leading order in $\lambda $,
and neglecting long-range correlations of the Majorana fermions containing higher order combinations of their density of states, we find two dominant contributions to the response $F(t)\approx F_{K}(t)+F_{H}(t)$.

Let us first consider the Raman response of the unperturbed Kitaev model, $ i F_K(t) = \langle 0| \operatorname{T}[ \mathcal{\hat{R}}_K(t)\mathcal{\hat{R}}_K(0)] |0 \rangle$, here $\operatorname{T}$ denotes the standard time-ordering. In terms of quasiparticle operators $\hat{a}_{\mathbf q}$ which diagonalize the flux-free Hamiltonian~(\ref{eq:HK}), the Raman operator is given by
\begin{eqnarray}
 	 \mathcal{\hat{R}}_K &=& \sum_{\mathbf q} \big\{ (h'_{\mathbf q} \cos 2\theta_{\mathbf q} - h''_{\mathbf q} \sin 2\theta_{\mathbf q} )
	\hat{a}^\dagger_{\mathbf q} \hat{a}^{\;}_{\mathbf q} \nonumber \\
	& &  + i ( h'_{\mathbf q} \sin 2\theta_{\mathbf q} + h''_{\mathbf q} \cos 2\theta_{\mathbf q} ) \hat{a}^\dagger_{\mathbf q} \hat{a}^{\dagger}_{-\mathbf q}
	+ {\rm h.c.} \big\},
\end{eqnarray}
where $h'_{\mathbf q}$ and $h''_{\mathbf q}$ denote the real and imaginary parts of $h_{\mathbf q} \equiv K \sum_{a=1}^3 (\hat{\bm\epsilon}_{\rm in}\cdot \mathbf d_a)(\hat{\bm\epsilon}_{\rm out}\cdot\mathbf d_a) e^{i \mathbf q\cdot\mathbf n_a}$ with $\mathbf n_0=(0,0)$ and $\mathbf {n}_1,\mathbf{n}_2$ defined in Fig.~\ref{fig:honeycomb}. Then
\begin{eqnarray}
\label{KitaevRamanContr}
	I_K(\omega) = 4\pi \sum_{\mathbf q} \delta(\omega - 4 |s_{\mathbf q}|) \left( {\rm Im}[ h^{\;}_{\mathbf q} s^*_{\mathbf q} ]/ |s_{\mathbf q}| \right)^2.
\end{eqnarray}

Next, we obtain the leading contribution to the Raman response due to the Heisenberg exchange $ i F_H(t) = \langle 0| \operatorname{T} [\mathcal{\hat{R}}_H(t)\mathcal{\hat{R}}_H(0) |0 \rangle$. A typical term in $\mathcal{\hat{R}}_H$, for example on a $z$ bond, contains spin operators $\propto \hat\sigma^x_{A,\mathbf r} \hat\sigma^x_{B,\mathbf r} + \hat\sigma^y_{A,\mathbf r} \hat\sigma^y_{B,\mathbf r}$. In terms of Majorana fermions, the spin operator, e.g. $\hat \sigma^a_{A, \mathbf r} = i c_{A, \mathbf r} ( \chi^{\;}_{\langle A,\mathbf r; B, \mathbf r + \mathbf n_a \rangle_a} + \chi^{\dagger}_{\langle A,\mathbf r; B, \mathbf r + \mathbf n_a \rangle_a} )$, creates a matter fermion $\hat{c}_i$ and changes the bond fermion number $\chi$ which corresponds to flipping the sign of two $Z_2$ fluxes on the plaquettes neighboring the bond (here the corresponding bond is of $x$ or $y$ type). The combined effect of these terms in $\mathcal{\hat R}_H$ is to insert four fluxes around the $z$-bond at site $\mathbf r$; see Fig.~\ref{fig:honeycomb}. The $Z_2$ fluxes have to be annihilated by the corresponding term in the other Raman operator in $iF_H(t)$ for a nonzero expectation value with respect to $|0\rangle$. Consequently, the Heisenberg Raman response $F_H$ can be decomposed into a sum over individual bonds $F_H(t) = \sum_{a=x,y,z}\sum_{ \mathbf r} F_{H, a}(\mathbf r, t)$. We focus on contributions from the $z$ bond (contributions from the $x$ and $y$ bonds can be obtained by symmetry).

The correlator $F_{H, z}(\mathbf r; t)$ contains two types of matrix elements, $\langle  \hat{\sigma}^x (t) \hat{\sigma}^x (t) \hat{\sigma}^x (0) \hat{\sigma}^x(0) \rangle$, and the off-diagonal ones $\langle  \hat{\sigma}^x (t) \hat{\sigma}^x (t) \hat{\sigma}^y (0) \hat{\sigma}^y(0) \rangle$. The corresponding correlators are denoted as $F^{xx}(t)$ and $F^{xy}(t)$. The former can be calculated without projection onto the physical states. However, the off-diagonal term conserves the flux sector, but {\em changes} the number of bond fermions, $\hat \chi$, thus one has to use the projectors in the calculation of $F^{xy}(t)$ \cite{Yao2009,Loss2011}, see supplementary material for details~\cite{suppl}.

The calculation of the Heisenberg part of the correlator can be cast as a local quantum quench, similar to the calculation of the dynamical spin correlation function in the Kitaev model~\cite{Baskaran2007,Knolle2014}. The Raman response can be expressed entirely in terms of  matter fermion operators acting in the ground-state flux sector, $|F_0\rangle$, subject to a dynamic local potential $\hat V$
\begin{eqnarray}\label{Fxx}
	& & F^{xx}_{H,z}(\mathbf r, t) = -i \langle M_0 | e^{i t \hat{\mathcal{H}}_0}\,e^{-i t (\hat{\mathcal{H}}_0 + \hat V_{\mathbf r}) } |M_0 \rangle, \nonumber \\
	& & F^{xy}_{H, z}(\mathbf r, t) = -i \langle M_0 | e^{i t \hat{\mathcal{H}}_0}\,e^{-i t (\hat{\mathcal{H}}_0 + \hat V_{\mathbf r})} c_{A, \mathbf r} c_{B, \mathbf r} | M_0\rangle.  \quad\quad
\end{eqnarray}
Here, the Hamiltonian $\hat{\mathcal{H}}_0 + \hat V_{\mathbf r}$ differs from $\hat{\mathcal{H}}_0$ in the sign of the Majorana hopping for the two $y$ bonds attached to sites $(A, \mathbf r)$ and $(B, \mathbf r)$. The locally perturbed Hamiltonian belongs to the sector with four extra fluxes shown in Fig.~\ref{fig:honeycomb}. The problem is now reduced to the one of a local quantum quench, where the ground state $|M_0\rangle$ of $\mathcal{\hat H}_0$ is time evolved with a different Hamiltonian $\hat{\mathcal{H}}_0 + \hat V_{\mathbf r}$. Note that in the calculation of dynamic spin-correlators in the Kitaev model, $\hat V_{\mathbf{r}}\propto \hat{f}_{\mathbf{r}}^{\dagger}\hat{f}_{\mathbf{r}}$ assumes the form of a local on-site potential~\cite{Knolle2014} which is switched on at $t=0$.
We have a four-flux rather than two-flux quench and the expression for $\hat V_{\mathbf{r}}$ in terms of complex bond-fermions is complicated. The correlators can be evaluated numerically using the Lehmann representation. To this end, we introduce a basis $|\lambda \rangle$ of many-body eigenstates of the Hamiltonian $\hat{\mathcal{H}}_0 + \hat V_{\mathbf r}$. We denote the corresponding energy as $E_\lambda$ and the ground-state energy of $\hat{\mathcal{H}}_0$ as $E_0$. We obtain
\begin{eqnarray}\label{heisenberg-intensity}
	& & I^{xx}_{H, z}(\omega) = 2\pi \sum_{\lambda} \delta\left( \omega - \Delta_\lambda \right) | \langle M_0 | \lambda \rangle|^2, \\
	& & I^{xy}_{H, z}(\omega) = 2\pi \sum_{\lambda} \delta\left( \omega - \Delta_\lambda \right) \langle M_0 | \lambda \rangle
	\langle \lambda | \hat c_{A,\mathbf{r}} \hat c_{B, \mathbf{r}} | M_0 \rangle, \nonumber
\end{eqnarray}
where $\Delta_\lambda=E_\lambda - E_0$. Note that non-zero contributions come only from excited states $|\lambda \rangle$ with the same parity as the ground state $|M_0\rangle$ of matter fermions.
We numerically calculate the dominant contributions $I_H^{\left[ 0\right] }(\omega), I_H^{\left[ 2\right] }(\omega)$ due to zero
and two-particle 
processes,
details are relegated to the supplementary material~\cite{suppl}.

\begin{figure}
\includegraphics[width=1.0\columnwidth]{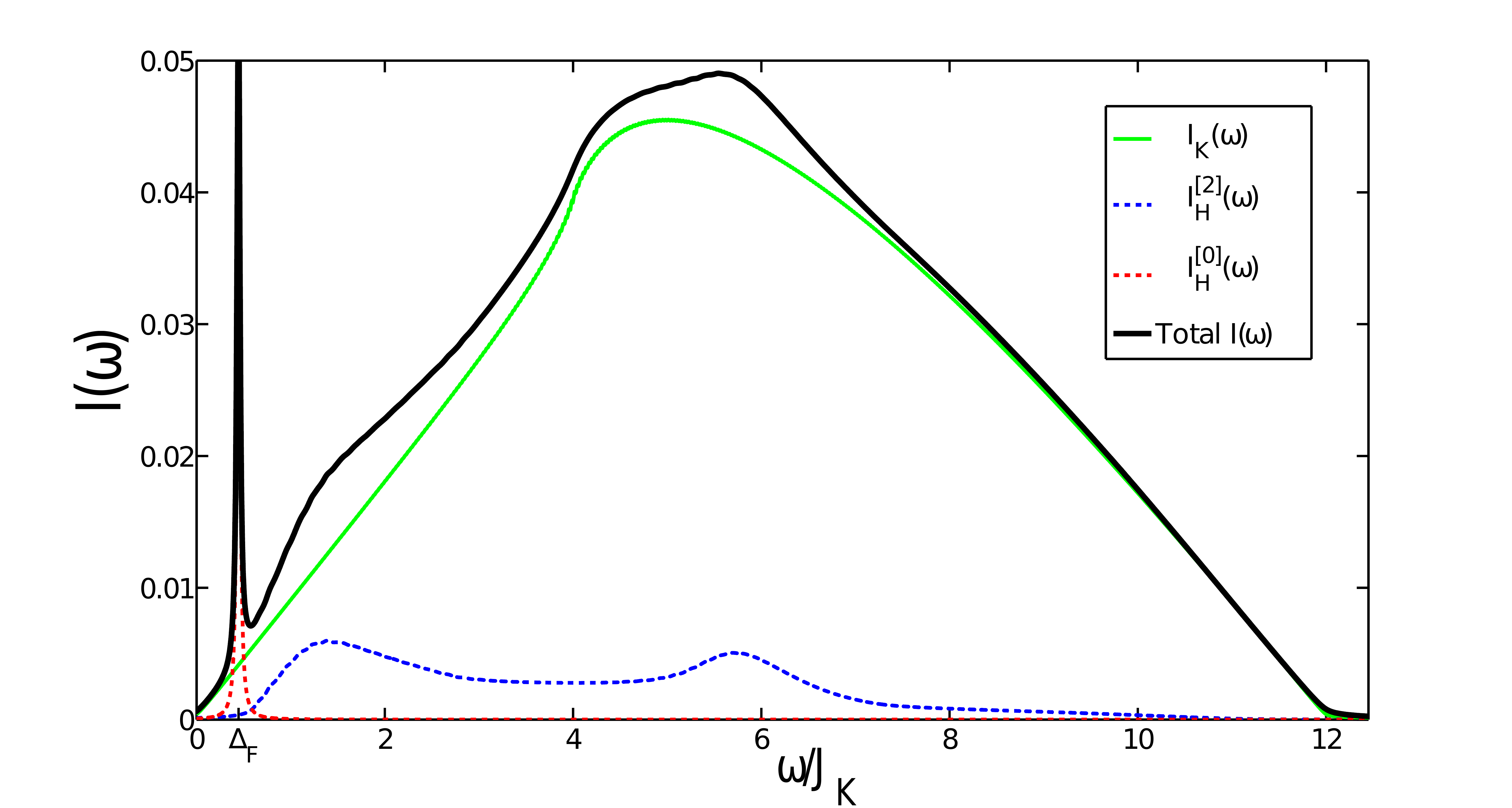}
\caption{\label{fig:spectrum} The Raman response $I(\omega)$ (black curve) and its various contributions (here $J_K=10 J_H$). The Kitaev contribution $I_K(\omega)$, shown in green, is independent of the photon polarization and shows characteristic features of the matter fermion density of states including the linear onset at low energies and the band-edge at $12 J_K$, note the additional factor of $2$ in Eq.(\ref{KitaevRamanContr}). The van Hove singularity at $2 J_K$ is seen as a small dip at $4J_K$ (a discontinuity of the derivative). The zero and two-particle responses, $I^{[0]}_H(\omega)$  and $I^{[2]}_H(\omega)$, of the Heisenberg Raman contribution are shown in blue and red (dashed) respectively. A $\delta$-function peak occurs at the four flux gap $\Delta_F = 0.446 J_K$, while the frequency dependence of the two-particle contribution reflects the local two-particle density of states in the presence of four fluxes.}
\end{figure}
{\it Results.} 
The Raman response, shown in Fig.~\ref{fig:spectrum}, is markedly different
from the known strongly polarization dependent behavior seen in the two-magnon response in antiferromagnetically ordered systems. In fact, the characteristic features of the weakly polarization-dependent response $I(\omega)$ can be related either to the flux, or the Majorana fermion sector: 
First, the polarization-independent Kitaev contribution $I_K(\omega)$ reflects the Majorana matter fermion density of states in the ground state flux sector. It has a linear onset at low energies, a sharp band-edge at $12J_K$, and a dip at $4J_K$ due to the van Hove singularity. 
Second, the Heisenberg contribution, which has a weak polarization dependence with a simple overall intensity change, is related to flux excitations, e.g.~$I_H(\omega)=0$ for $\omega<\Delta_F$.  A striking feature is a sharp peak at the energy of the four flux gap $\Delta_F=0.446 J_K$ originating from the zero-particle contribution (overlap between ground states), see Eq.~(\ref{heisenberg-intensity}). This is a clear signature of a pure flux excitation for the \textit{isotropic} gapless QSL ($J^x_K=J^y_K=J^z_K$). Note that normally sharp lines in Raman scattering are attributed to optical phonons which appear at different energy scales~\cite{devereaux07}. In addition, $I_H(\omega)$ has a broad response in energy reflecting the two-particle density of states of matter fermions propagating in the background of four inserted fluxes.

Our analysis of the Raman response relies on the stability of the Kitaev QSL with respect to addition of small Heisenberg exchanges (note that the latter is believed to be small in the proposed Kitaev model realisations in
Iridates). 
We expect that for small Heisenberg couplings the features that we find are robust, being only
somewhat renormalized by non-local fluctuations originating from the dynamics of the fluxes generated by the Heisenberg exchange (or disorder which is present in real materials). Crucially, there is a
window of parameters where the features that we find should be observable, thus making Raman scattering an important experimental tool for diagnosing Kitaev QSLs.

{\it Discussion.}
The calculation of the Heisenberg part $I_H(\omega)$ of the Raman response is equivalent to a non-equilibrium problem with a sudden insertion of four fluxes. 
The Raman vertex of the Kitaev model does not change the flux sector, but the integrability breaking contribution due to Heisenberg interactions does. The latter takes the form of a quantum quench which generates an unusual sharp $\delta$-function component in the response.

In general, for Kitaev type models we expect that the calculation of correlation functions $\langle \hat O (t) \hat O(0) \rangle$, whose operators $\hat O$ change the flux sector, can be mapped to a local quantum quench for Majorana fermions by exploiting selection rules and by eliminating flux degrees of freedom as pioneered for the spin correlation function in the original Kitaev model~\cite{Baskaran2007}.
This is true, for example, for the calculation of spin correlations in generalizations of the honeycomb model to higher dimensions~\cite{Mandal2009,Subhro2014,Kimchi2014,Hermanns2014,Takagi2014} (or possibly even to different classes~\cite{Yao2009,Qi2014,Vaezi2014}). The response will mainly be determined by the low energy matter fermions; e.g.~depending on the Fermi surface topology, a singular behavior may appear. Overall, while a QSL might be stable with respect to sufficiently weak integrability breaking interactions, the change in response functions can be remarkable, revealing basic properties of the underlying phase by connecting otherwise orthogonal sectors of the emergent gauge flux.

In conclusion, we have shown that Raman scattering renders visible both flux and Majorana fermion excitations potentially relevant to Iridates. It thus 
presents a valuable tool for diagnosing topological quantum states.

We thank N.~Shannon, S.~Bhattacharjee and especially J.~T.~Chalker for valuable discussions. J.K.~acknowledges support from the Studienstiftung des deutschen Volkes, the IMPRS Dynamical Processes in Atoms, Molecules and Solids, and DFG within GRK 1621. N.P.~acknowledges support from the  NSF grant DMR-1255544. J.~K.~is grateful for the hospitality of the Okinawa Institute of Science and Technology (OIST) where the final part of this work was completed. D.K.~ acknowledges EPSRC grant no.~EP/J017639/1.

\vspace{3cm}

\clearpage
\begin{widetext}
\section{Supplementary material}
Here, we provide details of our perturbation theory in small Heisenberg coupling. In addition, we present the calculation of $I_H(\omega)$ contribution to the Raman response. We also discuss the relevance of projection operators (to the physical subspace) for our calculations.

\subsection{Perturbation theory}
We derive the leading contributions to Eq.(\ref{RamanInteractionRep}) at small $K_H/K_K=J_H/J_K=\lambda \ll 1$. We explicitly take $\lambda$ out of the definition of the coupling constants such that $\mathcal{ H}=\mathcal{ H}_K+\lambda \mathcal{ H}_H$ and $\mathcal{R}=\mathcal{R}_K+\lambda \mathcal{R}_H$. The contributions of the Heisenberg Hamiltonian and Raman vertex diagonal in the flux sectors are absorbed into the  Kitaev exchange constants such that $J_{K} \to J_{K}(1+\lambda)$, $K_{K} \to K_{K}(1+\lambda)$. Hence every term in $\mathcal{ H}_H$ and $\mathcal{R}_H$ changes the flux sector when acting on the Kitaev ground state $|0\rangle$. 

The expression for the response reads (all operators are given in the
interaction representation, where the Heisenberg term is treated as
interaction)%
\begin{align}
i F(t)  & =\langle0|S^{\dagger}(t,-\infty)\mathcal{R}(t)S(t,0)\mathcal{R}(0)S(0,-\infty)|0\rangle,\\
S(t,t^{\prime})  & =\operatorname{T} \operatorname{exp} \left\{  -i \lambda \int_{t^{\prime}}^{t}%
\mathcal{ H}_{H}(\tau)d\tau\right\} \\
& =\operatorname{T}\left\{ 1 + \lambda \underbrace{(-i) \int_{t^{\prime}}^{t} d\tau \mathcal{ H}_{H}(\tau)}_{\equiv h^{(1)}_{H} (t,t^{\prime})} +
\lambda^2 \underbrace{\frac{(-i)^2}{2} \int_{t^{\prime}}^{t} d\tau_1 \int_{t^{\prime}}^{t} d\tau_2 \mathcal{ H}_{H}(\tau_1) \mathcal{ H}_{H}(\tau_2)}_{\equiv h^{(2)}_{H} (t,t^{\prime})} + \cdots \right\},
\end{align}
and we expand to leading order
\begin{align}\label{RamanSecondOrder}
i F(t)
\approx \langle0| \mathcal{R}_K(t) \mathcal{R}_K(0)|0\rangle + \lambda^2 \langle0| \mathcal{R}_H(t) \mathcal{R}_H(0)|0\rangle.
\end{align}

All terms linear in $\lambda$ vanish due to orthogonality of the flux sectors. The final two terms in Eq.~(\ref{RamanSecondOrder}) are the main contributions $F_K$ and $F_H$. Note that the contribution $F_H$ is a sum of purely local terms, see discussion in the next section, especially Eq.~(\ref{Ram1}).

The omitted $\lambda^2$ terms  involve non-local contributions. For example, four fluxes which are inserted by the term $\mathcal{R}_H$  can be locally annihilated by a suitable term in $\mathcal{ H}_H$ and the remaining matter degrees of freedom are non-local. Such multi-particle excitations are proportional to density-density fluctuations~\cite{tikhonov11} having higher orders of the matter fermion density of states and would contribute only a  correction which is broad in energy and in particular small at low energies~\cite{tikhonov11}. This is because in a Dirac system with only Fermi points instead of a proper Fermi surface, particle-hole excitations are suppressed at low energies due to the linearly vanishing density of states~\cite{Kotov2012}.

\subsection{Heisenberg contribution}
The vertex $\mathcal{R}_{{H}}  =  \sum_{\langle i j  \rangle_{\alpha}} \mathcal{R}_{{H}, \langle i j \rangle_{\alpha}}$ can be separated into three parts from the inequivalent bond directions
$\mathcal{R}_{{H}}  =   \sum_{\langle i j  \rangle_{z}} K_H \left( \epsilon_{\text{in}} {\mathbf d_{z}} \right) \left( \epsilon_{\text{out}} {\mathbf d_{z}} \right)
\left[  \sigma_i^{x}  \sigma_j^x +\sigma_i^{y}  \sigma_j^y \right] +
     \sum_{\langle i j  \rangle_{x}} K_H \left(\hat \epsilon_{\text{in}} {\mathbf d_{x}} \right) \left(\hat \epsilon_{\text{out}} {\mathbf d_{x}} \right)
\left[  \sigma_i^{z}  \sigma_j^z +\sigma_i^{y}  \sigma_j^y \right] +
 \sum_{\langle i j  \rangle_{y}} K_H \left(\hat \epsilon_{\text{in}} {\mathbf d_{y}} \right) \left(\hat \epsilon_{\text{out}} {\mathbf d_{y}} \right)
\left[  \sigma_i^{x}  \sigma_j^x +\sigma_i^{z}  \sigma_j^z \right]
$.
 Each of the terms $ \mathcal{R}_{{H}, \langle i j \rangle_{\alpha}}$  puts in four fluxes around the $\alpha$-bond at site $\mathbf{r}$, see Fig.\ref{fig:honeycomb}.
 The orthogonality of the flux sector greatly simplifies the calculation which turns out to be bond diagonal
 \begin{eqnarray}
\label{Ram1}
i F_{{H}} (t) = \langle \mathcal{R}_{{H}}(t) \mathcal{R}_{{H}} (0)\rangle
= \sum_{\langle i j\rangle_{\alpha}} \sum_{\langle k l\rangle_{\beta}} \langle \mathcal{R}_{{H}, \langle i j \rangle_{\alpha}} (t) \mathcal{R}_{{H}, \langle k l \rangle_{\beta}} (0) \rangle
=  \sum_{\langle i j\rangle_{\alpha}} \langle \mathcal{R}_{{H}, \langle i j \rangle_{\alpha}} (t) \mathcal{R}_{{H}, \langle i j \rangle_{\alpha}} (0)  \rangle.
\end{eqnarray}

We  concentrate on the Raman operator of a single $z$ bond,
 e.g. $\mathcal{R}_{{H}, \langle A\mathbf{r} B\mathbf{r} \rangle_{z}} \propto \sigma_{A\mathbf{r}}^x \sigma_{B\mathbf{r}}^x +\sigma_{A\mathbf{r}}^y \sigma_{B\mathbf{r}}^y$. The other contributions can be obtained  by cyclic permutation of the exchange constants $J_x,J_y,J_z$ and a rotation of the in- and out-going scattering angles by multiples of  $\frac{2\pi}{3}$.
The Raman intensity from the $z$-bond is given by
\begin{eqnarray}
\label{Ram2}
i F_{{H},z} (t) = && \lambda^2 \left[ K_K \left(\hat \epsilon_{\text{in}} {\mathbf d_{z}} \right) \left(\hat \epsilon_{\text{out}} {\mathbf d_{z}} \right) \right]^2 \langle
\left[ \sigma_{A\mathbf{r}}^x (t) \sigma_{B\mathbf{r}}^x (t) +\sigma_{A\mathbf{r}}^y(t) \sigma_{B\mathbf{r}}^y (t) \right]
\left[ \sigma_{A\mathbf{r}}^x (0) \sigma_{B\mathbf{r}}^x (0) +\sigma_{A\mathbf{r}}^y(0) \sigma_{B\mathbf{r}}^y (0) \right] \rangle
\end{eqnarray}
with two different types of matrix elements: spin component diagonal, e.g. $\langle \sigma^x(t) \sigma^x(t) \sigma^x(0) \sigma^x(0) \rangle$, and off-diagonal ones, e.g. $\langle \sigma^x (t)\sigma^x(t) \sigma^y(0) \sigma^y (0)\rangle$. The former operator can be calculated without the projection onto
the physical state because it neither changes the flux sector nor the bond fermion number.~\cite{Baskaran2007} The off-diagonal term conserves the flux sector but does change the bond fermion number $\chi$. Therefore, it is necessary to project the contribution from this term onto the physical state.\cite{Yao2009}  However, while the correlation function $\langle \sigma^x (t) \sigma^x(t) \sigma^y \sigma^y \rangle$ is non-zero, we find that its contribution to the Raman response is canceled by the $\langle \sigma^y (t) \sigma^y(t) \sigma^x \sigma^x \rangle$ term.

The calculation of the diagonal term
\begin{eqnarray}
\label{Ram3}
\langle \sigma_{A\mathbf{r}}^x (t) \sigma_{B\mathbf{r}}^x (t) \sigma_{A\mathbf{r}}^x (0) \sigma_{B\mathbf{r}}^x (0)\rangle  = &&
\langle e^{it{\mathcal{H}}_0 }  i c_{A\mathbf{r}} \left[ \chi_{\langle A\mathbf{r},B\mathbf{r}+\mathbf n_x \rangle_x} + \chi^{\dagger}_{\langle A\mathbf{r},B\mathbf{r}+\mathbf n_x \rangle_x} \right]
                        c_{B\mathbf{r}} \left[ \chi_{\langle A\mathbf{r}-\mathbf n_x ,B\mathbf{r}\rangle_x} - \chi^{\dagger}_{\langle A\mathbf{r}-\mathbf n_x, B\mathbf{r}\rangle_x} \right]  \times \\ \nonumber
&& e^{-it {\mathcal{H}}_0} i c_{A\mathbf{r}} \left[ \chi_{\langle A\mathbf{r},B\mathbf{r}+\mathbf n_x \rangle_x} + \chi^{\dagger}_{\langle A\mathbf{r},B\mathbf{r}+\mathbf n_x \rangle_x} \right]
                        c_{B\mathbf{r}} \left[ \chi_{\langle A\mathbf{r}-\mathbf n_x, B\mathbf{r}\rangle_x} - \chi^{\dagger}_{\langle A\mathbf{r}-\mathbf n_x, B\mathbf{r}\rangle_x} \right] \rangle
\end{eqnarray}
proceeds in a similar fashion as pioneered by Baskaran et al.~\cite{Baskaran2007} for the spin correlation function. The aim is to eliminate the bond fermions and to work entirely in the ground state flux sector. Recall that  the expectation value is taken over the ground state $|0\rangle= |F_0\rangle |M_0\rangle$ and we work in a gauge with $ \chi_i^{\dagger} \chi_i |F_0\rangle= |F_0\rangle $. We commute all bond operators to the right
\begin{eqnarray}
\label{ChiCommute}
\chi^{\dagger}_{\langle A\mathbf{r},B\mathbf{r}+\mathbf n_x \rangle_x} \chi^{\dagger}_{\langle A\mathbf{r}-\mathbf n_x,B\mathbf{r}\rangle_x} e^{-it {\mathcal{H}}_0} =
e^{-it {\mathcal{H}}_K \left[ \langle A\mathbf{r},B\mathbf{r}+\mathbf n_x \rangle_x, \langle A\mathbf{r}-\mathbf n_x,B\mathbf{r}\rangle_x\right] } \chi^{\dagger}_{\langle A\mathbf{r},B\mathbf{r}+\mathbf n_x \rangle_x} \chi^{\dagger}_{\langle A\mathbf{r}-\mathbf n_x, B\mathbf{r}\rangle_x}
\end{eqnarray}
 where ${\mathcal{H}}_K \left[ \langle A\mathbf{r},B\mathbf{r}+\mathbf n_x \rangle_x, \langle A\mathbf{r}-\mathbf n_x, B\mathbf{r}\rangle_x\right]$  is the Majorana hopping Hamiltonian with flipped link variables $u_{\langle A\mathbf{r},B\mathbf{r}+\mathbf n_x \rangle_x}=-1$ and $u_{\langle A\mathbf{r}-\mathbf n_x, B\mathbf{r}\rangle_x}=-1$ which corresponds to four fluxes around the $z$-bond. In the following we use the shorthand notation $\mathcal{ H}_K \left[ \langle A\mathbf{r},B\mathbf{r}+\mathbf n_x \rangle_x, \langle A\mathbf{r}-\mathbf n_x, B\mathbf{r}\rangle_x\right]=\mathcal{ H}_K \left[+\mathbf n_x, -\mathbf n_x \right]$.
We obtain an expression entirely in terms of matter fermions. The diagonal correlation function is
\begin{eqnarray}
\label{Ram4}
\langle \sigma_{A\mathbf{r}}^x (t) \sigma_{B\mathbf{r}}^x (t) \sigma_{A\mathbf{r}}^x (0) \sigma_{B\mathbf{r}}^x (0)\rangle  = &&
\langle M_0| e^{it  {\mathcal{H}}_0 } c_{A\mathbf{r}} c_{B\mathbf{r}} e^{-it  {\mathcal{H}}_K \left[+\mathbf n_x,-\mathbf n_x \right] } c_{B\mathbf{r}} c_{A\mathbf{r}} |M_0\rangle~.
\end{eqnarray}
This expression can be further simplified with the gauge equivalent
$c_{A\mathbf{r}} c_{B\mathbf{r}} e^{-it {\mathcal{H}}_K \left[+\mathbf n_x,-\mathbf n_x\right] } c_{B\mathbf{r}} c_{A\mathbf{r}} =
 e^{-it c_{A\mathbf{r}} c_{B\mathbf{r}}  {\mathcal{H}}_K \left[+\mathbf n_x,-\mathbf n_x\right] c_{B\mathbf{r}} c_{A\mathbf{r}}}
 =  e^{-it   {\mathcal{H}}_K \left[+\mathbf n_y,-\mathbf n_y \right] }
$
such that we recover the first line of Eq.~(\ref{Fxx})
\begin{eqnarray}
\label{Ram6}
\langle \sigma_{A\mathbf{r}}^x (t) \sigma_{B\mathbf{r}}^x (t) \sigma_{A\mathbf{r}}^x (0) \sigma_{B\mathbf{r}}^x (0)\rangle  = &&
\langle M_0| e^{it {\mathcal{H}}_0 } e^{-it  {\mathcal{H}}_K \left[+\mathbf n_y,-\mathbf n_y \right] } |M_0\rangle.
\end{eqnarray}
This is  a quantum quench in which the ground state $|M_0\rangle$ of $ {\mathcal{H}}_0$ is time evolved with the four flux Hamiltonian $ {\mathcal{H}}_K \left[+\mathbf n_y,-\mathbf n_y \right]= { \mathcal{H}}_0+{V}_{\mathbf{r}}$.

Next, we study the off-diagonal term $\langle \sigma^x (t)\sigma^x(t) \sigma^y(0) \sigma^y (0)\rangle$ for which it is necessary to include the projection operator
$\hat P|\Phi\rangle= \prod_{j} \frac{1+\hat D_j}{2}|\Phi\rangle=|\Phi_{\text{phys}}\rangle$
with $\hat D_j= b_j^xb_j^yb_j^z c_j$. Note that operators $\hat D_j$ commute with  the Hamiltonian ${\mathcal{H}}_K$ and all spin operators $\hat \sigma_i^a$. The projection can be factorized into~\cite{Yao2009}
\begin{eqnarray}
\label{Projection2}
\hat P= \hat P'\frac{1+\prod_j \hat D_j}{2} = \frac{\hat P'\left[ 1+(-1)^{N_{\chi}}(-1)^{N_{f}}\right] }{2}
\end{eqnarray}
where $\hat P'$ is the sum of all operators that change the bond fermion number in an inequivalent way. Here $N_{\chi}$ and $N_{f}$ are the total number of bond and matter fermions. We work with even total fermion number and use the fact that $\hat P$ commutes with all spin operators
such that
$\langle \hat P' \sigma_{A\mathbf{r}}^x (t) \sigma_{B\mathbf{r}}^x (t) \sigma_{A\mathbf{r}}^y (0) \sigma_{B\mathbf{r}}^y (0) \hat P' \rangle  =
\langle   \sigma_{A\mathbf{r}}^x (t) \sigma_{B\mathbf{r}}^x (t) \sigma_{A\mathbf{r}}^y (0) \sigma_{B\mathbf{r}}^y (0) \hat P' \rangle
$
which can be simplified in a similar fashion as before to
\begin{eqnarray}
\label{Ram7}
 && \langle \hat P \sigma_{A\mathbf{r}}^x (t) \sigma_{B\mathbf{r}}^x (t) \sigma_{A\mathbf{r}}^y (0) \sigma_{B\mathbf{r}}^y (0) \hat P \rangle   =
-\langle e^{it{\mathcal{H}}_0} e^{-it {\mathcal{H}}_K \left[+\mathbf{n_y},-\mathbf{n_y}\right] } \times \\ \nonumber
&&  \chi^{\dagger}_{\langle A\mathbf{r},B\mathbf{r}+\mathbf n_x \rangle_x}  \chi^{\dagger}_{\langle A\mathbf{r}-\mathbf n_x , B\mathbf{r}\rangle_x}  \left[ \chi_{\langle A\mathbf{r},B\mathbf{r}+\mathbf n_y \rangle_y} + \chi^{\dagger}_{\langle A\mathbf{r}, B\mathbf{r}+\mathbf n_y \rangle_y} \right]
                       \left[ \chi_{\langle A\mathbf{r}-\mathbf n_y,B\mathbf{r}\rangle_y} - \chi^{\dagger}_{\langle A\mathbf{r}-\mathbf n_y,B\mathbf{r}\rangle_y} \right] \hat P' \rangle.
\end{eqnarray}
In this expression only the part $\hat D_{A\mathbf{r}} \hat D_{B\mathbf{r}}$ of $\hat P'$  together with the product of $\chi$ operators does not change the bond fermion number.
On the one hand, the operator $\hat D_{A\mathbf{r}} \hat D_{B\mathbf{r}}$ eliminates the additional bond fermions but, on the other hand, it also introduces additional matter fermions. We recover the second  Eq.~(\ref{Fxx})
\begin{eqnarray}
\label{Ram6}
\langle \sigma_{A\mathbf{r}}^x (t) \sigma_{B\mathbf{r}}^x (t) \sigma_{A\mathbf{r}}^y (0) \sigma_{B\mathbf{r}}^y (0)\rangle  = &&
\langle M_0| e^{it  {\mathcal{H}}_0 } e^{-it {\mathcal{H}}_K \left[+\mathbf n_y, -\mathbf n_y \right] } c_{A\mathbf{r}} c_{B\mathbf{r}} |M_0\rangle.
\end{eqnarray}

\subsection{Leading few-particle contributions: zero- and two-particle response}

We insert a complete set of states $\sum_{\lambda} |\lambda \rangle \langle \lambda|$ into Eq.(\ref{Fxx}) with many-body eigenstates $|\lambda \rangle$ of the four flux Hamiltonian $ {\mathcal{H}}_K \left[+\mathbf n_y, -\mathbf n_y \right]={\mathcal{H}}_0+{{V}}_{\mathbf{r}}$. After Fourier transformation to the frequency domain, we obtain the expressions for the Raman intensity $ I_{H, \, z}^{xx} (\omega)$ and $I_{H,\, z}^{xy} (\omega)$ presented in Eq.\ref{heisenberg-intensity} of the main text.
In the sum over $\lambda$, in general, all multi-particle processes $b^{\dagger}_{\lambda} ... b^{\dagger}_{\lambda'} |M_F\rangle$ contribute with operators $b^{\dagger}_{\lambda}$ diagonalizing the four flux Hamiltonian of the matter sector with the ground state $|M_F\rangle$. In the following, we derive  formulas of the leading few-particle contributions. This approximation is expected to be very good because, due to the vanishing density of states, at low energies higher number particle processes are suppressed. For example, in the case of the pure Kitaev model at the isotropic point ($J_x=J_y=J_z$), already 98 \% of the total exact spin structure factor  were captured by single particle excitations.~\cite{Knolle2014}

In a short hand notation, let $b$ and $a$ be the eigenmodes  of (or operators that diagonalize) the system with extra fluxes and the one without extra fluxes, respectively, (written in terms of bond fermions $f=(c_A+ic_B)/2$ ) such that
\begin{eqnarray}
\label{BogNoFlux}
\begin{pmatrix}
X_0^* & Y_0^*\\
Y_0 & X_0
\end{pmatrix}
\begin{pmatrix}
f\\
f^{\dagger}
\end{pmatrix} =
\begin{pmatrix}
a\\
a^{\dagger}
\end{pmatrix}
 \  \
\text{and} \ \
\begin{pmatrix}
 X_F^* &  Y_F^*\\
 Y_F &  X_F
\end{pmatrix}
\begin{pmatrix}
f\\
f^{\dagger}
\end{pmatrix} =
\begin{pmatrix}
 b\\
 b^{\dagger}
\end{pmatrix}.
\end{eqnarray}
The two flux sectors can be related via
\begin{eqnarray}
\label{FluxNoFluxRelation}
\begin{pmatrix}
 \mathcal{X}^* & \mathcal{Y}^*\\
\mathcal{Y} & \mathcal{X}
\end{pmatrix}
\begin{pmatrix}
a\\
a^{\dagger}
\end{pmatrix} =
\begin{pmatrix}
 b\\
 b^{\dagger}
\end{pmatrix}
\  \
\text{with} \  \
\begin{pmatrix}
 \mathcal{X}^* & \mathcal{Y}^*\\
\mathcal{Y} & \mathcal{X}
\end{pmatrix}
=
\begin{pmatrix}
X_F^*  X_0^T+ Y_F^* Y_0^T & X_F^*  Y_0^{\dagger}+ Y_F^*  X_0^{\dagger}\\
Y_F  X_0^T+X_F  Y_0^T & Y_F  Y_0^{\dagger}+X_F  X_0^{\dagger}
\end{pmatrix}.
\end{eqnarray}
The ground state of the system with flux , $b |M_{F}\rangle =0$, can be obtained from the ground state of the flux free system, $a |M_0\rangle =0$, \cite{Ripka1985} as
\begin{eqnarray}
\label{GSrelation}
 |M_F\rangle = \text{det}\left( \mathcal{X}^{\dagger} \mathcal{X}\right)^{\frac{1}{4}} e^{-\frac{1}{2} \mathcal{F}_{ij} a_i^{\dagger} a_j^{\dagger}} |M_0\rangle
\ \ \text{with}  \  \ \mathcal{F}_{ij} = \left[ \mathcal{X}^{* -1}\right]_{il} \mathcal{V}^*_{lj}
\end{eqnarray}
which leads to the overlap $\langle M_F|M_0 \rangle  = \text{det}\left( \mathcal{X}^{\dagger} \mathcal{X}\right)^{\frac{1}{4}}$.

The first contribution to the sum of the Lehmann representation Eq.\ref{heisenberg-intensity} comes from the ground state overlaps ($|\lambda\rangle=|M_0\rangle$)
\begin{eqnarray}
\label{Lehmann}
I_{H, z}^{xx,[0]} (\omega) & =  &
 2 \pi \delta \left( \omega -\Delta_F \right) \sqrt{\text{det} \left( \mathcal{X}^{\dagger} \mathcal{X}  \right) } \\
I_{H, z}^{xy,[0]} (\omega) & = &
 i 2 \pi \delta \left( \omega -\Delta_F \right) \sqrt{\text{det} \left( \mathcal{X}^{\dagger} \mathcal{X}  \right) }
\left[ 1-2\sum_k |Y_{k0}|^2 -2 \sum_{j,k} Y_{0k}^{\dagger} \mathcal{F}^*_{kj} X _{j0}^* \right].
\end{eqnarray}
We have numerically studied systems up to $62\times 62$ unit cells (7688 spins). The zero particle response is nonzero if the overlap between the ground states $|\langle M_0|M_F\rangle|$ is nonzero.  It turns out that this is indeed the case since both have the same parity. In addition, there is no Anderson orthogonality catastrophe for such a Dirac system.~\cite{Knolle2014} Hence, we have derived a $\delta$-function contribution at the four flux gap $\Delta_F$ to the total Raman response.

The next signal from single particle contributions $b^{\dagger}_{\lambda} |M_F\rangle$ are zero because of opposite parity to the zero flux ground state $|M_0\rangle$.
Only even numbers of particles contribute. Frequency dependence first arises from two particle contributions  $b^{\dagger}_{\lambda}b^{\dagger}_{\lambda'} |M_F\rangle$
\begin{eqnarray}
\label{Lehmann1}
I_{H, z}^{xx,[2]} (\omega) & =  &
2 \pi \sum_{\lambda,\lambda'} \delta \left( \omega -\left[ E_{\lambda}+E_{\lambda'}- E_0\right]  \right) |\langle M_0|b^{\dagger}_{\lambda} b^{\dagger}_{\lambda'}|M_F \rangle|^2 \\ \nonumber
I_{H, z}^{xy,[2]} (\omega) & =  &
2 \pi \sum_{\lambda,\lambda'} \delta \left( \omega -\left[ E_{\lambda}+E_{\lambda'}- E_0\right]  \right) \langle M_0|b^{\dagger}_{\lambda} b^{\dagger}_{\lambda'}|M_F \rangle \langle M_F | b_{\lambda'} b_{\lambda} c_{A0} c_{B0} | M_0\rangle .
\end{eqnarray}
The matrix elements are given at the end. It turns out that the off-diagonal contributions $I_{H, z}^{xy} (\omega)$, which involve the projection operator, are purely imaginary. They are exactly canceled by the complex conjugate  $I_{H, z}^{yx} (\omega)$ and do not contribute to the Raman response.

The matrix elements for the two particle contribution, Eq.\ref{Lehmann1}, are
\begin{eqnarray}
\label{G2}
 G^{[2]}_{\lambda \lambda'} = \langle M_0|b^{\dagger}_{\lambda} b^{\dagger}_{\lambda'}|M_F \rangle  =  \text{det}\left( \mathcal{X}^{\dagger} \mathcal{X}\right)^{\frac{1}{4}} \left\lbrace   \mathcal{Y}_{\lambda l} \mathcal{X}^T_{l \lambda'} + \mathcal{Y}_{\lambda l} \mathcal{F}_{l k} \mathcal{Y}^T_{k \lambda'}\right\rbrace
\end{eqnarray}
and
\begin{eqnarray}
\label{G4}
G^{[4]}_{\lambda \lambda'} & = & \langle M_0|c_{B0} c_{A0} b^{\dagger}_{\lambda} b^{\dagger}_{\lambda'}|M_F \rangle  = -i G^{[2]}_{\lambda \lambda'} +  g^{[4]}_{\lambda \lambda'} +  \tilde g^{[4]}_{\lambda \lambda'}  \  \ \text{with}  \\
g^{[4]}_{\lambda \lambda'} & = & 2 i \text{det}\left( \mathcal{X}^{\dagger} \mathcal{X}\right)^{\frac{1}{4}} \left\lbrace
\mathcal{X}_{\lambda i} X_{i0} \mathcal{X}_{\lambda' j} Y_{j0} - \mathcal{X}_{\lambda i} Y_{i0} \mathcal{X}_{\lambda' j} X_{j0} + Y_{i0}^{*} Y_{0i} \mathcal{Y}_{\lambda l} \mathcal{X}^T_{l \lambda'}  \right\rbrace \\
\tilde g^{[4]}_{\lambda \lambda'} & = & 2 i \text{det}\left( \mathcal{X}^{\dagger} \mathcal{X}\right)^{\frac{1}{4}}  \{
\mathcal{Y}_{\lambda l} \mathcal{X}_{l \lambda'}^T  X_{0j}^T \mathcal{F}_{ji} Y_{i0} + \mathcal{Y}_{\lambda l} \mathcal{F}_{li} X^T_{0i} \mathcal{X}_{\lambda' j} Y_{j0} + \mathcal{Y}_{\lambda l} \mathcal{F}_{li}^T Y_{i0} \mathcal{X}_{\lambda' j} X_{j0} + \\ \nonumber
& & \mathcal{X}_{\lambda i} X_{i0} \mathcal{Y}_{\lambda' l} \mathcal{F}_{lj} Y_{j0} -
 \mathcal{X}_{\lambda i} Y_{i0} \mathcal{Y}_{\lambda' l} \mathcal{F}_{lj} X_{j0} -
\mathcal{Y}_{\lambda l} \mathcal{F}_{lk} \mathcal{Y}_{k \lambda'}^T Y_{0j}^T Y_{j0}^* \}.
\end{eqnarray}
\end{widetext}
\end{document}